\documentstyle[prl,aps,epsf,multicols]{revtex}

\begin{document}
\title{Scalar Field Dark Matter}
\author{Tonatiuh Matos\thanks{%
E-mail: tmatos@fis.cinvestav.mx}, F. Siddhartha Guzm\'{a}n\thanks{%
E-mail: siddh@fis.cinvestav.mx} and L. Arturo Ure\~{n}a-L\'{o}pez\thanks{%
E-mail: lurena@fis.cinvestav.mx}}
\address{Departamento de F\'{\i}sica,\\
Centro de Investigaci\'on y de Estudios Avanzados del IPN,\\
A.P. 14-740, 07000 M\'exico D.F., MEXICO.\\
}
\author{Dar\'{\i}o N\'{u}\~{n}ez\thanks{
E-mail: nunez@nuclecu.unam.mx}}
\address{Instituto de Ciencias Nucleares,\\
Universidad Nacional Aut\'onoma de M\'exico,\\
A.P. 70-543, 04510 M\'exico D.F., MEXICO}
\date{\today}
\date{\today}
\maketitle

\begin{abstract}
This work is a review of the last results of research on the Scalar Field
Dark Matter model of the Universe at cosmological and at galactic level. We
present the complete solution to the scalar field cosmological scenario in
which the dark matter is modeled by a scalar field $\Phi$ with the scalar
potential $V(\Phi )=V_{0}\left[ \cosh {(\lambda \, \sqrt{\kappa _{0}}\Phi)}-1%
\right]$ and the dark energy is modeled by a scalar field $\Psi$, endowed
with the scalar potential $\tilde{V}(\Psi )= \tilde{V_{0}}\left[\sinh {%
(\alpha \,\sqrt{\kappa _{0}}\Psi )}\right]^{\beta }$, which together compose
the $95\%$ of the total matter energy in the Universe. The model presents
successfully deals with the up to date cosmological observations, and is a
good candidate to treat the dark matter problem at the galactic level.
\end{abstract}


\begin{multicols}{2}
\narrowtext

\section{Introduction}

There is no doubt that we are living exiting times in physics. Certainly,
the last results of the observations of the Universe will conduce to a
qualitative new knowledge of Nature. One question which singles out is that
of finding out which is the nature of the matter energy composing the
Universe. It is amazing that after so much effort dedicated to such a
question, what is the Universe composed of?, it has not been possible to
give a conclusive answer. From the latest observations, we do know that
about 95\% of matter in the Universe is of non baryonic nature. The old
belief that matter in Cosmos is made of quarks, leptons and gauge bosons is
being abandoned due to the recent observations and the inconsistencies which
spring out of this assumption \cite{schram}. We think there is strong
evidence on the existence of an exotic non baryonic sort of matter which
dominates the structure of the Universe, but its nature is until now a
puzzle.

In this work we pretend to summarize the results of one proposal for the
nature of the matter in the Universe, namely, the Scalar Field Dark Matter
model (SFDM). This model has had relative success at cosmological level as
well as at galactic scale. In this work we pretend to put both models
together and explain which is the possible connection between them. 

The hypothesis that the scalar field is the dark matter is not new. For
example, Denhen has proposed that the Higgs scalar field could be the dark
matter in galaxies \cite{dehnen1,dehnen2}. Other authors has suggested that
the halo of galaxies are boson stars, etc. Maybe the most popular hypothesis
was the axions fields as dark matter. Nevertheless, this hypothesis show
some problems of consistency\cite{seidel,seidel98}. In our opinion, it was
not possible to do a model of the Universe till the accelerated expansion of
it was established. Most of the material contained in this work has been
separately published elsewhere, the main goal of this work is to write it
all together and with a unifying description. First, let us give a brief
introduction of the Dark Matter problem.

The existence of dark matter in the Universe has been firmly established by
astronomical observations at very different length-scales, ranging from
single galaxies, to clusters of galaxies, up to cosmological scale (see for
example \cite{falta}). A large fraction of the mass needed to produce the
observed dynamical effects in all these very different systems is not seen.
At the galactic scale, the problem is clearly posed: The measurements of
rotation curves (tangential velocities of objects) in spiral galaxies show
that the coplanar orbital motion of gas in the outer parts of these galaxies
keeps a more or less constant velocity up to several luminous radii \cite
{persic,begeman}, forming a nearly radii independent curve in the outer
parts of the rotational curves profile; a motion which does not correspond
to the one due to the gravitational effects of the observed matter
distribution, hence there must be present some type of dark matter causing
the observed motion. The flat profile of the rotational curves is the
clearest indication of the presence of more than the observed matter. It is
usually assumed that the dark matter in galaxies has an almost spherical
distribution which decays like $1/r^{2}$. With this distribution of some
kind of matter it is possible to fit the rotational curves of galaxies quite
well \cite{begeman}. Nevertheless, the main question of the dark matter
problem remains: Which is the nature of the dark matter in the Universe? The
problem is not easy to solve, it is not sufficient to find out an exotic
particle which could exist in galaxies in the low energy regime of some
theory. It is necessary to show as well, that this particle distributes in a
very similar manner in all these galaxies, and finally, to give even some
reason for its existence in galaxies.

Recent observations of the luminosity-redshift relation of Ia Supernovae
suggest that distant galaxies are moving slower than predicted by Hubble's
law impliying an accelerated expansion of the Universe \cite{snia}. These
observations open the possibility to the existence of an energy component in
the Universe with a negative equation of state, $\omega <0$, being $p=\omega
\rho $, called dark energy. This result is very important because it
separates the components of the Universe into gravitational (also called the
``matter component'' of the Universe) and antigravitational (repulsive, the
dark energy). The dark energy would be the currently dominant component in
the Universe and its ratio relative to the whole energy would be $%
\Omega_{DE}\sim 70\%$. The most simple model for this dark energy is the
cosmological constant ($\Lambda$), for which $\omega=-1$.

Observations in galaxy clusters and dynamical measurements of the mass in
galaxies indicate that the matter component of the Universe is $%
\Omega_{M}\sim 30\%.$ But this component of the Universe decomposes itself
in baryons, neutrinos, etc. and cold dark matter, which is the responsible
of the formation of structure in the Universe. Observations indicate that
stars and dust (baryons) represent something like $0.3\%$ of the whole
matter of the Universe. The new measurements of the neutrino mass indicate
that neutrinos contribute with a same quantity like dust. In other words,
say $\Omega_{M}=\Omega _{b}+\Omega _{\nu }+ \cdot \cdot \cdot+\Omega
_{DM}\sim 0.05+\Omega _{DM}$, where $\Omega _{DM}$ represents the dark
matter part of the matter contributions which has a value of $%
\Omega_{DM}\sim 0.25$. The value of the amount of baryonic matter ($\sim 5\%$%
) is in concordance with the bounds imposed by nucleosynthesis (see for
example \cite{schram}). These observations are in very good agreement with
the preferred value $\Omega _{0}\sim 1.$

The standard cosmological model then considers a flat Universe ($\Omega
_{\Lambda }+\Omega _{M}\approx 1$) full with $95\%$ of unknown matter which
is of great importance at cosmological level. Moreover, it seems to be the
most successful model fitting current cosmological observations \cite
{triangle}.

The SFDM model \cite{varun,siddh,urena,luis2} consists in to assume that the
dark matter and the dark energy are of scalar field nature. A particular
model we have considered in the last years is the following. For the dark
energy we adopt a quintessence field ${\Psi}$\ with a $\sinh$ like scalar
potential. In \cite{luis}, it was showed that the potential

\begin{eqnarray}
\tilde{V}(\Psi ) &=&\tilde{V_{0}}\left[ \sinh {(\alpha \,\sqrt{\kappa _{0}}
\Psi )}\right] ^{\beta }  \label{sinh} \\
&=&\left\{ 
\begin{array}{cc}
\tilde{V_{0}}\left( \alpha \,\sqrt{\kappa _{0}} \Psi \right) ^{\beta } & 
|\alpha \,\sqrt{\kappa _{0}} \Psi |\ll 1 \\ 
\left( \tilde{V_{0}}/2^{\beta }\right) \exp {\left( \alpha \beta \,\sqrt{%
\kappa _{0}} \Psi \right) } & |\alpha \,\sqrt{\kappa _{0}} \Psi |\gg 1
\end{array}
\right. ;  \nonumber
\end{eqnarray}

\noindent is a reliable model for the dark energy, because of its asymptotic
behaviors. Nevertheless, the main point of the SFDM model is to suppose that
the dark matter is a scalar field ${\Phi }$\ endowed with the potential \cite
{luis2}

\begin{eqnarray}
V(\Phi ) &=&V_{0}\left[ \cosh {(\lambda \,\sqrt{\kappa _{0}} \Phi )}-1\right]
\label{cosh} \\
&=&\left\{ 
\begin{array}{cc}
\frac{1}{2} m^2_\Phi \Phi^2 + \frac{1}{24} \lambda^2 m^2_\Phi \kappa_0 \Phi^4
& |\lambda \,\sqrt{\kappa _{0}} \Phi |\ll 1 \\ 
\left( V_{0}/2\right) \exp {\left( \lambda \,\sqrt{\kappa _{0}} \Phi \right)}
& |\lambda \,\sqrt{\kappa _{0}} \Phi |\gg 1
\end{array}
\right. .  \nonumber
\end{eqnarray}

\noindent The mass of the scalar field $\Phi$ is defined as $%
m_{\Phi}^{2}=V^{\prime \prime }|_{\Phi =0}=\lambda ^{2}\kappa _{0}V_{0}$. In
this case, we are dealing with a massive scalar field.

The main results of the model presented in this work are: 1) the fine tuning
and the cosmic coincidence problems are ameliorated for both dark matter and
dark energy and the models agrees with astronomical observations. 2) The
model predicts a suppression of the Mass Power Spectrum for small scales
having a wave number $k>k_{min,\Phi }$, where $k_{min,\Phi}\simeq 4.5\,h\,$
Mpc${}^{-1}$ for $\lambda \simeq 20.3$. This last fact could help to explain
the dearth of dwarf galaxies and the smoothness of galaxy core halos. 3)
From this, all parameters of the scalar dark matter potential are completely
determined. 4) The dark matter consists of an ultra-light particle, whose
mass is $m_{\Phi }\simeq 1.1\times 10^{-23}\,{\rm eV}$ and all the success
of the standard cold dark matter model is recovered. 5) If the scale of
renormalization of the model is of order of the Planck Mass, then the scalar
field $\Phi $ can be a reliable model for dark matter in galaxies. 6) The
predicted scattering cross section fits the value required for
self-interacting dark matter. 7) Studying a spherically symmetric
fluctuation of the scalar field $\Phi $ in cosmos we show that it could be
the halo dark matter in galaxies. 8) The local space-time of the fluctuation
of the scalar field $\Phi $ contains a three dimensional space-like
hypersurface with surplus of angle. 9) We also present a model for the dark
matter in the halos of spiral galaxies, we obtain that the effective energy
density goes like $1/(r^{2}+K^{2})$ and 10) the resulting circular velocity
profile of tests particles is in good agreement with the observed one in
spiral galaxies.

All these facts lead us to consider that the scalar field is a very good
candidate for being the dark matter of the Universe.

\section{Cosmological Scalar Field Solutions}

In this section, we give all the solutions to the model at the cosmological
scale and focus our attention in the scalar dark matter. Since current
observations of CMBR anisotropy by {\small BOOMERANG} and {\small MAXIMA} 
\cite{boom} suggest a flat Universe, we use as ansatz the flat
Friedmann-Robertson-Walker (FRW) metric

\begin{equation}
ds^{2}=-dt^{2}+a^{2}(t)\left[ dr^{2}+r^{2} d\Omega^2 \right] ,
\end{equation}

\noindent where $a$ is the scale factor ($a=1$ today) and we have set $c=1$.
The components of the Universe are baryons, radiation, three species of
light neutrinos, etc., and two minimally coupled and homogenous scalar
fields $\Phi $ and $\Psi $, which represent the dark matter and the dark
energy, respectively. The evolution equations for this Universe read

\begin{eqnarray}
H^{2} \equiv \left( \frac{\dot{a}}{a} \right)^2 &=&\frac{\kappa _{0}}{3}
\left( \rho +\rho _{\Phi }+\rho _{\Psi }\right)  \label{fried} \\
\ddot{\Phi}+3H\dot{\Phi}+\frac{dV(\Phi )}{d\Phi } &=&0  \label{cphi} \\
\ddot{\Psi}+3H\dot{\Psi}+\frac{d\tilde{V}(\Psi )}{d\Psi } &=&0  \label{cpsi}
\\
{\dot{\rho}}+3H\left( \rho +p\right) &=&0,  \label{fluid}
\end{eqnarray}

\noindent being $\kappa _{0} \equiv 8\pi G$ and $\rho$ ($p$) is the energy
density (pressure) of radiation, plus baryons, plus neutrinos, etc. The
scalar energy densities (pressures) are $\rho _{\Phi }=\frac{1}{2}\dot{\Phi}
^{2}+V(\Phi )$ ($p_\Phi =\frac{1}{2}\dot{\Phi}^{2}-V(\Phi )$) and $%
\rho_{\Psi }=\frac{1}{2}\dot{\Psi}^{2}+\tilde{V}(\Psi )$ ($p_\Psi =\frac{1}{%
2 }\dot{\Psi}^{2}-\tilde{V}(\Psi )$). Here overdots denote derivative with
respect to the cosmological time $t$.

\subsection{Radiation Dominated Era (RD)}

We start the evolution of the Universe at the end of inflation, $i.e.$ in
the radiation dominated (RD) era. The initial conditions are set such that $%
\left( \rho_{i\Phi },\,\rho_{i\Psi }\right) \leq \rho_{i\gamma }$. Let us
begin with the dark energy. For the potential (\ref{sinh}) an exact solution
in the presence of nonrelativistic matter can be found\cite{luis,chimen} and
the parameters of the potential are given by

\begin{eqnarray}
\alpha &=&\frac{-3\omega _{\Psi }}{2\sqrt{3(1+\omega _{\Psi })}},  \nonumber
\\
\beta &=&\frac{2\,(1+\omega _{\Psi })}{\omega _{\Psi }},  \label{tracpsi} \\
\kappa_0 \tilde{V_0} &=& \frac{3(1-\omega_\Psi)}{2} \left( \frac{\Omega_{oM}%
}{\Omega_{o\Psi}} \right)^{\frac{1+\omega_\Psi}{\omega_\Psi}} \Omega_{o\Psi}
H^2_0  \nonumber
\end{eqnarray}

\noindent with $\Omega _{o\Psi }$ and $\Omega _{oCDM}$ the current values of
dark energy and dark matter, respectively; and $-0.9 \leq \omega _{\Psi}\leq
-0.6$ the range for the current equation of state. With these values for the
parameters ($\alpha ,\,\beta <0$) the solution for the dark energy ($\Psi $)
becomes a tracker one and is only reached until a matter dominated epoch.
The scalar field $\Psi $ would begin to dominate the expansion of the
Universe after matter domination. Before this, at the radiation dominated
epoch, the scalar energy density $\rho _{\Psi }$ is frozen, strongly
subdominant and of the same order than today \cite{luis}. Then the dark
energy contribution can be neglected during this epoch.

Now we study the behavior of the dark matter. For the potential (\ref{cosh})
we begin the evolution with large and negative values of $\Phi$, when the
potential behaves as an exponential one. It is found that the exponential
potential makes the scalar field $\Phi$ mimic the dominant energy density,
that is, $\rho_\Phi = \rho_{i\Phi} a^{-4}$. The ratio of $\rho_\Phi$ to the
total energy density is \cite{chimen,ferr}

\begin{equation}
\frac{\rho _{\Phi }}{\rho _{\gamma }+\rho _{\Phi }}=\frac{4}{\lambda ^{2}}.
\label{rad}
\end{equation}

\noindent This solution is self-adjusting and it helps to avoid the fine
tuning problem of matter, too. Here appears one restriction due to
nucleosynthesis \cite{ferr} acting on the parameter $\lambda> \sqrt{24}$.
Once the potential (\ref{cosh}) reaches its polynomial behavior, $\Phi$
oscillates so fast around the minimum of the potential that the Universe is
only able to feel the average values of the energy density and pressure in a
scalar oscillation. Both $\Phi$ and $<\omega_\Phi>$ go down to zero and $%
<\rho_\Phi>$ scales as non-relativistic matter \cite{ford}. If we would like
the scalar field $\Phi $ to act as cold dark matter, in order to recover all
the successful features of the standard model, we need first derive a
relation between the parameters ($V_0$, $\lambda$). The required relation
reads \cite{luis2}

\begin{equation}
\kappa_0 V_0 \simeq \frac{1.7}{3} \left( \lambda^2 -4\right)^3 \left(\frac{%
\Omega_{0CDM}}{\Omega_{o\gamma}} \right)^3 \Omega_{0CDM} H^2_0.
\label{ratio}
\end{equation}

\noindent Notice that $V_0$ depends on both current amounts of dark matter
and radiation (including light neutrinos) and that we can choose $\lambda$
to be the only free parameter of potential (\ref{cosh}). Since now, we can
be sure that $\rho_{\Phi}=\rho_{CDM}$ and that we will recover the standard
cold dark matter evolution.

\subsection{Matter Dominated Era (MD) and Scalar Field $\Psi$ Dominated Era (%
$\Psi$D)}

During this time, the scalar field $\Phi$ continues oscillating and behaving
as nonrelativistic matter and there is a matter dominated era just like that
of the standard model. A short after matter completely dominates the
evolution of the Universe, the scalar field $\Psi$ reaches its tracker
solution and it begins to be an important component. Lately, the scalar
field $\Psi$ becomes the dominant componente of the Universe and the scalar
potential (\ref{sinh}) is effectively an exponential one \cite{urena,luis}.
Thus, the scalar field $\Psi$ drives the Universe into a power-law
inflationary stage ($a \sim t^p$, $p>1$). This solution is distinguishable
from a cosmological constant one \cite{luis}.

A complete numerical solution for the dimensionless density parameters $%
\Omega $'s are shown in Fig. 1 up to date. The results agree with the
solutions found in this section. It can be seen that eq. (\ref{ratio}) makes
the scalar field $\Phi$ behave quite similar to the standard cold dark
matter model once the scalar oscillations start and the required
contributions of dark matter and dark energy are the observed ones \cite
{luis2,luis,luis3}. We recovered the standard cosmological evolution and
then we can see that potentials (\ref{sinh},\ref{cosh}) are reliable models
of dark energy and dark matter in the Universe.

\begin{figure*}[tbp]
\centerline{ \epsfysize=5cm \epsfbox{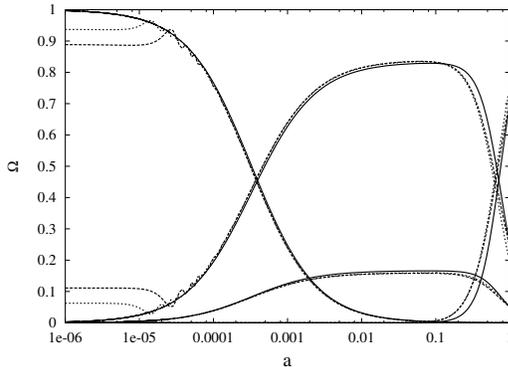}}
\caption{Evolution of the dimensionless density parameters {\it vs} the
scale factor $a$ with $\Omega_{oM} = 0.30$: $\Lambda$CDM (solid-curves) and $%
\Psi\Phi DM$ for two values of $\protect\lambda=6$ (dashed-curves), $\protect%
\lambda=8$ (dotted-curves). The equation of state for the dark energy is $%
\protect\omega_\Psi = -0.8$.}
\label{fig:Omegac}
\end{figure*}

\subsection{Scalar Power Spectrum for $\Phi$ dark matter}

In this section, we analyze the perturbations of the space due to the
presence of the scalar fields $\Phi,\Psi$. First, we consider a linear
perturbation of the space given by $h_{ij}$. We will work in the synchronous
gauge formalism, where the line element is $ds^{2}=a^{2}[-d\tau
^{2}+(\delta_{ij}+h_{ij})dx^{i}dx^{j}]$. We must add the perturbed equations
for the scalar fields $\Phi(\tau) \rightarrow \Phi(\tau) + \phi(k,\tau)$ and 
$\Psi(\tau) \rightarrow \Psi(\tau) + \psi(k,\tau)$\cite{ferr,stein}

\begin{eqnarray}
\ddot{\phi} + 2 {\cal H}\dot{\phi} + k^2 \phi + a^2 V^{\prime \prime} \phi + 
\frac{1}{2} \dot{\Phi} \dot{h} &=& 0  \label{pphi} \\
\ddot{\psi} + 2 {\cal H} \dot{\psi} + k^2 \psi + a^2 \tilde{V}^{\prime
\prime} \psi + \frac{1}{2} \dot{\Psi} \dot{h} &=& 0 .  \label{ppsi}
\end{eqnarray}

\noindent to the linearly perturbed Einstein equations in $k$-space ($\vec{k}
= k \hat{k}$) (see \cite{ma}). Here, overdots are derivatives with respect
to the conformal time $\tau$ and primes are derivatives with respect of the
unperturbed scalar fields $\Phi$ and $\Psi$, respectively.

It is known that scalar perturbations can only grow if the $k^{2}$-term in
eqs. (\ref{pphi},\ref{ppsi}) is subdominant with respect to the second
derivative of the scalar potential, that is, if $k<a \sqrt{V^{\prime \prime}}
$ \cite{ma2}. According to the solution given above for potential (\ref{cosh}%
), $k_{\Phi }=a \sqrt{V^{\prime \prime }}$ has a minimum value given by \cite
{luis3}

\begin{equation}
k_{min,\Phi} = m_\Phi a^{\ast} \simeq 1.3 \, \lambda \sqrt{\lambda^2 -4} 
\frac{\Omega_{0CDM}}{\sqrt{\Omega_{o\gamma}}} H_0 ,  \label{kmin}
\end{equation}

\noindent $a^\ast$ being the scale factor at the time when the scalar field $%
\Phi$ starts to oscillate coherently around the minimum of the scalar
potential~(\ref{cosh}). For $\lambda \geq 5$, we have that $k_{min,\Phi}
\geq 0.375 \,$ Mpc${}^{-1}$. Then, it can be assured that there are no
scalar perturbations for $k > m_\Phi$, that is, bigger than $k_\Phi$ today.
These $k$ correspond to scales smaller than $1.2$ kpc (here $\lambda=5$).
They must have been completely erased. Besides, modes which $m_\Phi > k >
k_{min,\Phi}$ must have been damped during certain periods of time. From
this, we conclude that the scalar power spectrum of $\Phi$ will be damped
for $k > k_{min,\Phi}$ with respect to the standard case. Therefore, the
Jeans length must be \cite{luis3}

\begin{equation}
L_J (a) = 2 \pi \, k^{-1}_{min,\Phi},  \label{jl}
\end{equation}

\noindent and it is a universal constant because it is completely determined
by the mass of the scalar field particle. $L_J$ is not only proportional to
the quantity $(\kappa_0 V_0)^{-1/2}$ (recalling that $m^2_\Phi=\lambda^2
\kappa_0 V_0$), but also the time when scalar oscillations start
(represented by $a^\ast$) is important.

On the other hand, the wave number for the dark energy $k_{\Psi }=a \sqrt{%
\tilde{V}^{\prime \prime }}$ is always out of the Hubble horizon, then only
structure at larger scales than $H^{-1}$ can be formed by the scalar
fluctuations $\psi$. Instead of a minimum, there is a maximum $%
k_{max,\Psi}\sim 10^{-3} \,$ Mpc${}^{-1}$. All scalar perturbations of the
dark energy which $k>10^{-3} \,$ Mpc${}^{-1}$ must have been completely
erased. Perturbations with $k\leq 10^{-3} \,$ Mpc${}^{-1}$ have started to
grow only recently. For a more detailed analysis of the dark energy
fluctuations, see \cite{ma2,brax}.

\begin{figure*}[tbp]
\centerline{ \epsfysize=5cm \epsfbox{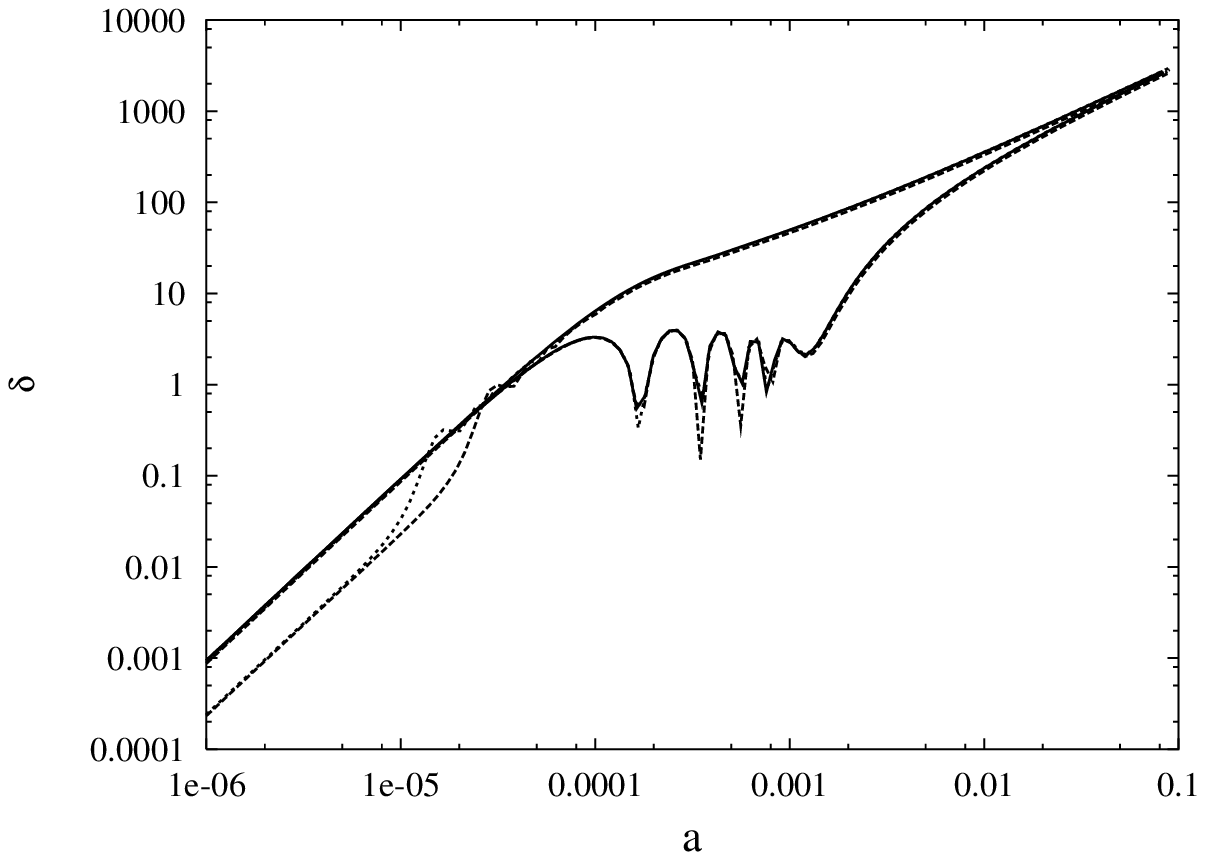}}
\centerline{ \epsfysize=5cm \epsfbox{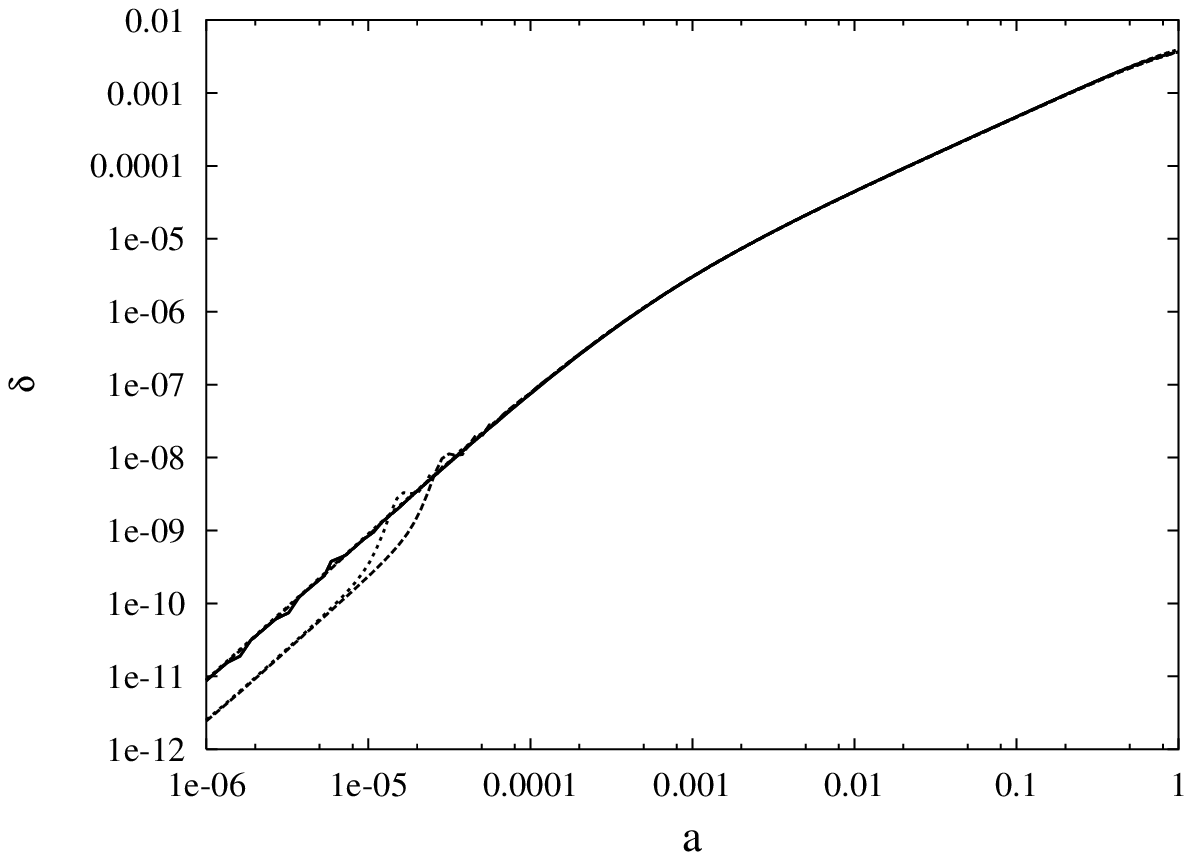}}
\caption{ Evolution of the density contrasts for baryons $\protect\delta_b$,
standard cold dark matter $\protect\delta_{CDM}$ and scalar dark matter $%
\protect\delta_\Phi$ {\it vs} the scale factor $a$ taking $\Omega_{oM} =
0.30 $ for the models given in fig.~(\ref{fig:Omegac}). The modes shown are $%
k=1.0 \times 10^{-5} \,$ Mpc${}^{-1}$ (left) and $k=0.1 \,$ Mpc${}^{-1}$
(right).}
\label{fig:dphi}
\end{figure*}

In Fig.~\ref{fig:dphi}, a numerical evolution of the density contrasts $\delta \equiv
\delta \rho / \rho$ is shown compared with the standard CDM case\cite{luis2}%
. The evolution was done using an amended version of {\small CMBFAST} \cite
{seljak}. Due to its oscillations around the minimum, the scalar field $\Phi$
changes to a complete standard CDM and so do its perturbations. All the
standard growing behavior for modes $k < k_{min,\Phi}$ is recovered and
preserved until today by potential (\ref{cosh}).

\begin{figure*}[tbp]
\centerline{ \epsfysize=5cm \epsfbox{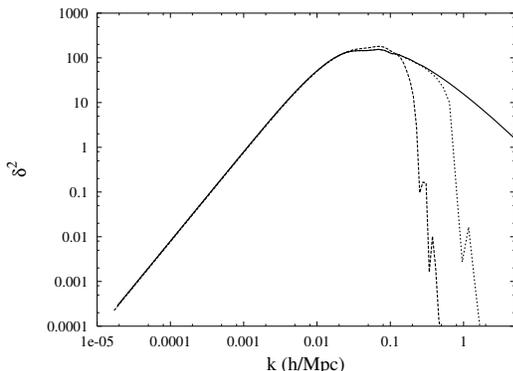}}
\caption{Power spectrum at a redshift $z=50$: $\Lambda$CDM (solid-curve),
and $\Phi $CDM with $\protect\lambda = 5$ (dashed-curve) and $\protect%
\lambda = 10$ (dotted-curve). The normalization is arbitrary.}
\label{fig:tf}
\end{figure*}

In Fig.~\ref{fig:tf} we can see $\delta ^{2}$ at a redshift $z=50 $ from a
complete numerical evolution using the amended version of {\small CMBFAST}\cite{luis3}.We also observe a sharp cut-off in the processed power spectrum at small scales when compared to the standard case, as it was argued above. This
suppression could explain the smooth cores of dark halos in galaxies and a
less number of dwarf galaxies \cite{kamion}.

\noindent The mass power spectrum is related to the CDM case by the
semi-analytical relation (see \cite{hu})

\begin{equation}
P_\Phi(k) \simeq \left( \frac{\cos{x^3}}{1+x^8} \right)^2 P_{CDM}(k),
\end{equation}

\noindent but using $x=(k/k_{min,\Phi})$ with $k_{min,\Phi}$ being the wave
number associated to the Jeans lenght (\ref{jl}). If we take a cut-off of
the mass power spectrum at $k=4.5 \, h \, {\rm Mpc}^{-1}$ \cite{kamion}, we
can fix the value of parameter $\lambda$. Using eq. (\ref{kmin}), we find
that \cite{luis3}

\begin{eqnarray}
\lambda &\simeq& 20.3,  \nonumber \\
V_0 &\simeq& \left(3.0\times 10^{-27}\,M_{Pl}\simeq 36.5\, {\rm eV}\right)
^{4}, \\
m_\Phi &\simeq& 9.1\times 10^{-52}\,M_{Pl}\simeq 1.1\times 10^{-23}\, {\rm eV%
}.  \nonumber
\end{eqnarray}

\noindent where $M_{Pl}=1.22 \times 10^{19} \, {\rm GeV}$ is the Plank mass.
All parameters of potential (\ref{cosh}) are now completely determined and
we have the right cut-off in the mass power spectrum.


\section{Scalar dark matter and Planck scale physics}

At galactic scale, numerical simulations show some discrepancies between
dark-matter predictions and observations \cite{steinC}. Dark matter
simulations show cuspy halos of galaxies with an excess of small scale
structure, while observations suggest a constant halo core density \cite
{avila} and a small number of subgalactic objects \cite{avila2}. There have
been some proposals for resolving the dark matter crisis (see for example 
\cite{steinC,avila2}). A promising model is that of a self-interacting dark
matter \cite{stein2}. This proposal considers that dark matter particles
have an interaction characterized by a scattering cross section by mass of
the particles given by \cite{avila,stein2}

\begin{equation}
\frac{\sigma_{2 \rightarrow 2}}{m} = 10^{-25}-10^{-23} cm^2 \, GeV^{-1}.
\label{sigma}
\end{equation}

\noindent This self-interaction provides shallow cores of galaxies and a
minimum scale of structure formation, that it must also be noticed as a
cut-off in the Mass Power Spectrum \cite{kamion,avila2}.

Because of the presence of the scalar field potential~(\ref{cosh}), there
must be an important self-interaction among the scalar particles. That means
that this scalar field dark matter model belongs to the so called group of
self-interacting dark matter models mentioned above, and it should be
characterized by the scattering cross section (\ref{sigma}).

Let us calculate $\sigma_{2\rightarrow 2}/m_{\Phi }$ for the scalar
potential (\ref{cosh}). The scalar potential (\ref{cosh}) can be written as
a series of even powers in $\Phi $, $V=\sum _{n=1}^{\infty }V_0\lambda
^{2n}\kappa _{0}^{n}\Phi^{2n}$. Working on 4 dimensions, it is commonly
believed that only $\Phi ^{4}$ and lower order theories are renormalizable.
But, following the important work \cite{halpern}, if we consider that there
is only one intrinsic scale $\Lambda $ in the theory, we conclude that there
is a momentum cut-off and that we can have an effective $\Phi^{4}$ theory
which depends upon all couplings in the theory. In addition, it was recently
demonstrated \cite{bran} that scalar exponential-like potentials of the form
(we have used the notation of potential (\ref{cosh}) and for example,
parameter $\mu $ in eqs. (7-8) in ref. \cite{bran} is $\mu ^{-1}=\lambda 
\sqrt{\kappa _{0}}$)) \cite{luis4}

\begin{equation}
U_\Lambda (\Phi) = M^4 \exp{\left( - \frac{\lambda^2 \kappa_0 \Lambda^2}{32
\pi^2} \right)} \exp{\left( \pm \lambda \sqrt{\kappa_0} \Phi \right)}
\label{renpot}
\end{equation}

\noindent are non-perturbative solutions of the exact renormalization group
equation in the Local Potential Approximation (LPA). Here $M$ and $\lambda $
are free parameters of the potential and $\Lambda$ is the scale of
renormalization. Being our potential for scalar dark matter a $\cosh $-like
potential (non-polynomial), it is then a solution to the renormalization
group equations in the (LPA), too. Comparing eqs. (\ref{cosh}, \ref{renpot}%
), we can identify $V_0=M^4 \exp{\left[ -\left(\lambda^2 \kappa_0 \Lambda^2
\right)/ (32 \pi^2) \right]}$. We see that an additional free parameter
appears, the scale of renormalization $\Lambda$. From this, we can assume
that potential (\ref{cosh}) is renormalizable with only one intrinsic scale: 
$\Lambda$.

Following the procedure shown in \cite{halpern2} for a potential with even
powers of the dimensionless scalar field $\phi$, $U(\phi)=\sum^{%
\infty}_{n=1} u_{2n} \phi^{2n}$, the cross section for $2 \rightarrow 2$
scattering in the center-of-mass frame is \cite{luis4}

\begin{equation}
\sigma_{2 \rightarrow 2} = \frac{M^8 \kappa^4_0 \lambda^8}{16 \pi E^2},
\label{sigma1}
\end{equation}

\noindent where $E$ is the total energy. Thus, we arrive to a real effective 
$\phi^4$ theory with a coupling $g=M^4 \kappa^2_0 \lambda^4$, leading to a
different cross section for the self-interacting scalar field than that
obtained from polynomial models \cite{bento}. Near the threshold $E \simeq
4m^2$, and taking the values of $\lambda$, $m_\Phi$ and the expected result (%
\ref{sigma}), we find $M= (6.7 \pm 1.9) \times 10^2$ TeV and $\Lambda \simeq
(1.93 \pm 0.01) \, M_{Pl} \simeq 2.3 \times 10^{19}$ GeV. The scale of
renormalization is of order of the Planck Mass. All parameters are
completely fixed now \cite{luis4}.


\section{Scalar Field Dark Matter in Galaxies}

In this section we explore whether a scalar field can fluctuate along the
history of the Universe and thus forming concentrations of scalar field
density. This idea was first explored in the early 90's for a Bose gas made
of Higgs bosons in the newtonian regime \cite{dehnen1,dehnen2,evag}.
Nevertheless, we assume that the halo of a galaxy is a fluctuation of
cosmological scalar dark matter and study the consequences for the
space-time background at this scale starting directly from a relativistic
point of view \cite{asnach}. Such relativistic approach for weakly
gravitating systems has its origins in the Projective Unified Field Theory 
\cite{schmut1,schmut2}, where it appears a scalar field as a consequence of
the projection to four dimensions.

The region of the galaxy we are interested in is that which goes from the
limits of the luminous matter (including the stars far away of the center of
the galaxy) over the limits of the halo. In this region measurements
indicate that circular velocity of stars is almost independent of the radii
at which they are located within the equatorial plane \cite{vera}.
Observational data show that the galaxies are composed by almost 90\% of
dark matter. Nevertheless the halo contains a larger amount of dark matter,
because otherwise the observed dynamics of particles in the halo is not
consistent with the predictions of Newtonian theory, which explains well the
dynamics of the luminous sector of the galaxy. So we can suppose that
luminous matter does not contribute in a very important way to the total
energy density of the halo of the galaxy at least in the mentioned region,
instead the scalar matter will be the main contributor to it. As a first
approximation we can neglect the baryonic matter contribution to the total
energy density of the halo of the galaxy. On the other hand, the exact
symmetry of the halo is stills unknown, but it is reasonable to suppose that
the halo is symmetric with respect to the rotation axis of the galaxy.
Furthermore, the rotation of the galaxy does not affect the motion of test
particles around the galaxy, dragging effects in the halo of the galaxy
should be too small to affect the tests particles (stars) traveling around
the galaxy. Hence, we can consider a time reversal symmetry of the
space-time. We consider that the dark matter will be the main contributor to
the dynamics, and then we will treat the observed luminous matter as a test
fluid. The line-element of such space-time, given in the Papapetrou form is 
\cite{kramer}:

\begin{equation}
ds^{2}=-e^{2\,\psi }(dt+\omega \,d\varphi )^{2}+e^{-2\psi }[e^{2\gamma
}(d\rho ^{2}+dz^{2})+\mu ^{2}d\varphi ^{2}],  \label{eq:ele}
\end{equation}

\noindent where $\psi ,\omega ,\gamma $, and $\mu $, are functions of $(\rho
,z)$.

We will derive the geodesic equations in the equatorial plane, that is for $%
z=\dot{z}=0$, where dot stands for the derivative with respect to the proper
time $\tau $. The Lagrangian for a test particle travelling on the static
space time ($\omega =0$) described by (\ref{eq:ele}) is given by:

\begin{equation}
2{\cal L}=-e^{2\,\psi} \dot{t}^{2}+e^{-2\psi } [e^{2\gamma }\, (\dot{\rho}%
^{2}+\dot{z}^{2})+\mu ^{2}\,\dot{\varphi}^{2}],  \label{eq:lag}
\end{equation}

In order to have stable circular motion, which is the motion we are
interested in, we have to satisfy three conditions:

i) $\dot{\rho}=0$, and

ii)${\frac{{\partial V(\rho)}}{{\partial\,\rho}}}=0$, where $%
V(\rho)=-e^{2(\psi-\gamma)}\,[E\,\dot{t}-L\,\dot{\varphi}-1] $.

iii)${\frac{{\partial^2 V(\rho)}}{{\partial\,\rho^2}}}|_{extr}>0$, in order
to have a minimum.

Under these conditions, we find the form of the line element in the
equatorial plane has to be \cite{nuestro}

\begin{equation}
ds^{2}=-\left( {\frac{{\mu }}{{\mu _{0}}}}\right) ^{2l}\,dt^{2}+\left( {%
\frac{{\mu }}{{\mu _{0}}}}\right) ^{-2l}\,[e^{2\,\gamma }\,d\rho ^{2}+\mu
^{2}\,d{\varphi }^{2}].  \label{rcmetric}
\end{equation}

\noindent where $l=(v^{\varphi})^2/(1+(v^{\varphi})^2)$, being $v^{\varphi}$
the tangential velocity of a test particle travelling on the equatorial
plane of the background space-time. Notice that this type of space time
definitely can not be asymptotically flat.

In what follows we study the circular trajectories of a test particle on the
equatorial plane taking the space-time (\ref{rcmetric}) as the background.
The motion equation of a test particle in such space-time can be derived
from the Lagrangian (\ref{eq:lag}), the angular momentum per unit of mass is

\begin{equation}
\frac{\mu ^{2}}{f}\frac{d\varphi }{d\tau }=B,  \label{B}
\end{equation}

\noindent where $f=e^{2\psi }=\left( {\mu /\mu _{0}}\right) {^{2l}}$ and the
total energy of the test particle reads

\begin{equation}
f\ c^{2}\frac{d\ t}{d\tau }=A,  \label{A}
\end{equation}

\noindent with $\tau $ is the proper time of the test particle. An observer
falling freely into the galaxy, with coordinates $(\rho ,\zeta ,\phi ,t)$,
will have a line element given by

\begin{eqnarray}
ds^{2} &=&\left\{ \frac{1}{f\ c^{2}}[e^{2\gamma }({\dot{\rho}}^{2}+{\dot{%
\zeta}}^{2})+\mu ^{2}{\dot{\varphi}}^{2}]-f\right\} c^{2}dt^{2}  \nonumber \\
&=&\left( \frac{v^{2}}{c^{2}}-f\right) c^{2}dt^{2}  \label{line} \\
&=&-c^{2}d\tau ^{2}.  \nonumber
\end{eqnarray}

The velocity ${\bf v}$ given by $v^{a}=(\dot{\rho},\dot{\zeta},\dot{\phi}) $%
, is the three-velocity of the test particle, where a dot means derivative
with respect to $t$, the time measured by the free falling observer. The
squared velocity $(v^{\varphi})^{2}$ is

\begin{equation}
v^{2}=g_{ab}v^{a}v^{b}=\frac{e^{2\gamma }}{f}(\dot{\rho}^{2}+\dot{\zeta}%
^{2})+\frac{\mu ^{2}}{f}\dot{\varphi}^{2},  \label{vel}
\end{equation}

\noindent where $a,b=1,2,3$. Using equation (\ref{line}) we obtain an
expression for $dt/d\tau $, by substituting it into (\ref{A}), we get

\begin{equation}
A^{2}=\frac{c^{4}f^{2}}{f-\frac{v^{2}}{c^{2}}}.  \label{Ac}
\end{equation}

For the axisymmetric scalar field dark matter halo, an exact solution of the
field equations reads in Boyer-Lindquist coordinates \cite{siddh}:

\begin{eqnarray}
ds^{2} &=& -f_{0}c^{2}\left( r^{2}+b^{2}\sin ^{2}{\theta }\right) dt^{2} \nonumber \\
&& + \frac{1+b^{2}\cos ^{2}{\theta }/r^{2}}{f_{0}}\left( \frac{dr^{2}}{1+b^{2}/r^{2}}+r^{2}d\theta ^{2}\right) \nonumber \\
&& +\frac{r^{2}+b^{2}\sin ^{2}{\theta }}{f_{0}}d\varphi ^{2}  \label{eq:metricBL}
\end{eqnarray}

\noindent and the effective energy density $\mu _{DM}$ is half the value of
the scalar potential:

\begin{eqnarray}
\mu _{DM}&=&\frac{1}{2}V(\Phi )=-\frac{2f_{0}r_{0}}{\kappa_{0}(r^{2}+b^{2}\sin ^{2}{\theta )}}, \nonumber \\
V(\Phi) &=& -\frac{l}{\kappa_0(1- l)} exp\left[-2\sqrt{\frac{\kappa_0}{l}}(\Phi - \Phi_0)\right]  \label{trash}
\end{eqnarray}

\noindent being $\mu = \sqrt{r^2 + b^2}\sin \theta$ and $\zeta = r \cos
\theta$ in (\ref{rcmetric}).

The geodesic equations of the metric (\ref{eq:metricBL}) in the equatorial
plane read

\begin{eqnarray}
\frac{d^{2}D}{d\tau ^{2}}-D\left( \frac{d\varphi }{d\tau }\right)^{2}+f_{0}^{2}c^{2}D\left( \frac{dt}{d\tau }\right) ^{2}&=&0,\nonumber \\
 \frac{d\varphi }{d\tau }&=&\frac{B}{D^{2}},\nonumber \\
\frac{dt}{d\tau }&=&\frac{A}{c^{2}f_{0}^{2}D^{2}}  \label{geo}
\end{eqnarray}

\noindent where $\tau $ is the proper time of the test particle and $D=\int {%
ds}=\sqrt{(r^{2}+b^{2})/f_{0}}$ is the proper distance of the test particle
at the equator from the galactic center. Observe that for a circular
trajectory it follows

\begin{equation}
\frac{d\varphi }{dt}=f_{0}c=c^{2}f_{0}^{2}\frac{B}{A}
\end{equation}

\noindent Moreover $B=cD$ along the whole galaxy. The first of equations (%
\ref{geo}) is the second Newton's law for particles travelling onto the
scalar field background. We can interpret

\begin{eqnarray}
\frac{d^{2}D}{d\tau ^{2}}&=&D\left( \frac{d\varphi }{d\tau }\right)
^{2}-f_{0}^{2}c^{2}D\left( \frac{dt}{d\tau }\right) ^{2} \nonumber \\
&=&\frac{B^{2}}{D^{3}}-\frac{A^{2}}{c^{2}D^{3}}=\frac{c^{2}}{D}-\frac{A^{2}}{c^{2}f_{0}^{2}D^{3}}
\label{chr5}
\end{eqnarray}

\noindent as the force due to the scalar field background, i.e. $F_{\Phi
}=f_{0}^{2}c^{2}/D-A^{2}/\left( c^{2}f_{0}^{2}D^{3}\right) $. Using the
expression for $A$ given in equation (\ref{Ac}), we can write down this
force in terms of $v^{\varphi}$

\begin{equation}
F_{\Phi }=-\frac{(v^{\varphi})^{2}}{D\left(
f_{0}^{2}D^{2}-(v^{\varphi})^{2}/c^{2}\right) }
\end{equation}

\noindent which corresponding gravitational potential due to the prsence of
the scalar field` is

\begin{equation}
V_{\Phi }=\frac{1}{2}c^{2}\ln \left( f_{0}^{2}-\frac{1}{D^{2}}\frac{v^{2}}{%
c^{2}}\right)  \label{Vp}
\end{equation}

This potential corresponds to non stable trajectories. Nevertheless, $%
v^{2}/c^{2}\simeq 10^{-6}$, and $f_{0}\simeq 0.01$ kpc${}^{-1}$, this means
that potential $V_{\Phi }$ is almost constant for $D\sim 0.3$ kpc, which
corresponds to the region where the solution is valid. At the other hand, we
know that the luminous matter is completely Newtonian (possesses small
velocities, provokes weak gravitational field and is dust). The Newtonian
force due to the luminous matter is then given by $%
F_{L}=GM(D)/D^{2}=v_{L}^{2}/D=B_{L}^{2}/D^{3}$, where $v_{L}$ is the
circular velocity of the test particle due to the contribution of the
luminous matter and $B_{L}=B_{L}(D)$ is its corresponding angular momentum
per unit of mass. The total force acting on the test particle is then $%
F=F_{L}+F_{\Phi }.$ For circular trajectories $d^{2}D/d\tau ^{2}=F=0$, then

\begin{equation}
F=\frac{B_{L}^{2}}{D^{3}}-\frac{A^{2}}{c^{2}f_{0}^{2}D^{3}}+\frac{c^{2}}{D}=0
\label{chr6}
\end{equation}

The corresponding potential is then $V=V_{L}+V_{\Phi }$, but for the regions
within $D \leq 0.3$ kpc, potential $V$ is dominated by the behavior of $%
V_{L} $, which contains stable circular trayectories, like in a galaxy. For
regions with $D\sim 0.3$ kpc the solution is not valid any more due to the
approximations we have carried out.

Using (\ref{Ac}) and (\ref{chr6}) we obtain an expression for $B$ in terms
of $(v^{\varphi})^{2}$,

\begin{eqnarray}
B_{L}^{2} &=&\frac{(v^{\varphi})^{2}D^{2}}{f_{0}^{2}D^{2}- \frac{%
(v^{\varphi})^{2}}{c^{2}}}  \nonumber \\
&\sim &(v^{\varphi})^{2}\frac{1}{f_{0}^{2}},  \label{Bc}
\end{eqnarray}

\noindent since $(v^{\varphi})^{2} \ll c^{2}$. Now using (\ref{Bc}) one
concludes that for our solution (\ref{eq:metricBL}) $(v^{%
\varphi})^{2}=f_{0}^{2}B_{L}^{2}$, $i.e.$

\begin{equation}
v_{DM}=f_{0}B_{L},  \label{Ma}
\end{equation}

\noindent where we call $v^{\varphi} \rightarrow v_{DM}$ the circular
velocity due to the dark matter.

The observed luminous matter in a galaxy behaves in accord to Newtonian
dynamics to a good approximation, so that its angular momentum per unit mass
will be $B_{L}=v_{L}\times D$, where $v_{L}$ is the contribution of the
luminous matter, and $D$ is the interval from the metric as written in
equation~(\ref{eq:metricBL}) with $D=\sqrt{(r^{2}+b^{2})/f_{0}}$. Since the
expression for $v_{DM}$ represents the velocity of test particles due to the
presence of the scalar field; we get:

\begin{equation}
v_{DM}=f_{0}v_{L}\sqrt{(r^{2}+b^{2})/f_{0}},  \label{eq:vdmb}
\end{equation}

\noindent Noting that the total kinetic energy is the sum of the individual
contributions, i.e., $mv_{C}^{2}/2=mv_{L}^{2}/2+mv_{gas}^{2}/2+mv_{DM}^{2}/2$%
, we arrive at the final form of the velocity along circular trajectories in
the equatorial plane of the galaxy:

\begin{equation}
v_{C}^{2}=v_{L}^{2}(1+f_{0}(r^{2}+b^{2}))+v_{gas}^{2},  \label{eq:vctot}
\end{equation}

\noindent where the constants $f_{0}$ and $b$ will be parameters to adjust
to the observed RC. The total luminous mass at a distance $r$ from the
center of the galaxy will be $M_{L}(r)=(M/L)\times L(r)$, i.e.:

\begin{equation}  \label{eq:vlum}
v_{L}^2(r) = \frac{GM_L(r)}{r}.
\end{equation}

\noindent Combining equations~(\ref{eq:vctot}) and~(\ref{eq:vlum}), and
including the 21 cm data from gas, we are led to:

\begin{figure*}
\label{fig:fits}
\centerline{ \epsfxsize=9cm \epsfbox{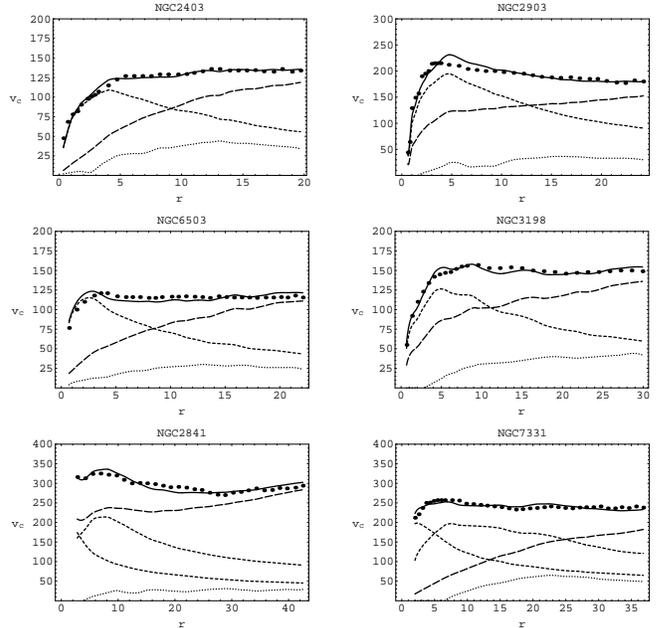}}
\caption{Curves of $v_C$ (km/s) vs $r$ (kpc) for the observational data
(dots) and the fits from equation (\ref{eq:modeltofit1}) (solid). Also shown
are the individual contributions from luminous matter (short dashed), the
scalar field (long dashed) and gas (dotted). In the case of NGC 2841 and NGC
7331, the lower short dashed curve represents the contribution from the
bulge of these galaxies.}
\end{figure*}

\begin{equation}
v_{C}^{2}(r)=\frac{GM_{L}(r)}{r}(1+f_{0}(r^{2}+b^{2}))+v_{gas}^{2}.
\label{eq:modeltofit1}
\end{equation}

The main results are shown in Fig. 4 and in Table 1 (see also 
\cite{rmaa}), in which there are
shown the observational RC (for simplicity we have omitted the error bars)
as well as the best fit parameters for the generated curves using equation~(%
\ref{eq:modeltofit1}). Shown are also the individual contributions from
luminous matter, gas and the scalar field. It can be noted that the
agreement is quite good (within 5\% in all cases), which could have been
expected since there are three parameters to be adjusted.

The remaining two parameters, $f_{0}$ and $b$, serve only to determine
completely the metric, and they do not have a direct physical interpretation
other than as part of the scalar field energy density.

\subsection{A particular case: spherical symmetry}

Analogously, assuming\ that the halo has spherically symmetry (a particular
case of the previous symmetry) the space-time with the {\it flat curve
condition}, it is easy to find the (see \cite{nuestro,sph}) becomes in
general

\begin{equation}
ds^{2}=-B_{0}r^{m}dt^{2}+A(r)dr^{2}+r^{2}d\theta ^{2}+r^{2}\sin ^{2}\theta
d\varphi ^{2}  \label{metrica}
\end{equation}

\noindent with $m=2(v^{\varphi })^{2}$ preserving the meaning of $v^{\varphi
}$. This result is not surprising. Remember that the Newtonian potential $%
\psi _{N}$ is defined as $g_{00}=-\exp (2\psi _{N})=-1-2\psi _{N}-\cdots $.
On the other side, the observed rotational curve profile in the dark matter
dominated region is such that the rotational velocity $v^{\varphi }$ of the
stars is constant, the force is then given by $F=-(v^{\varphi })^{2}/r$,
which respective Newtonian potential is $\psi =(v^{\varphi })^{2}\ln (r)$.
If we now read the Newtonian potential from the metric (\ref{metrica}), we
just obtain the same result. Metric (\ref{metrica}) is then the metric of
the general relativistic version of a matter distribution, which test
particles move in constant rotational curves. Function $A(r)$ will be
determined by the kind of substance we are supposing the dark matter is made
of.\newline

However, the case we are putting forward in this moments corresponds to the
case of a stress-energy tensor corresponding to a scalar field, for which,
when assumed the equivalent potential found for the axial case (\ref{trash})

\begin{equation}
V(\Phi) = -\frac{m}{\kappa_0 (2 -m)}\exp\left[-2\sqrt{\frac{\kappa_0}{m}}%
(\Phi - \Phi_0)\right]
\end{equation}

\noindent the interesting line element reads:

\begin{equation}
ds^{2}=-B_{0}r^{m}dt^{2}+dr^{2}+\frac{4}{4-m^{2}}r^{2}\left[ d\theta
^{2}+\sin ^{2}\theta d\varphi ^{2}\right]  \label{main}
\end{equation}

\noindent for which the two dimensional hypersurface area is $4\pi
r^{2}\times 4/(4-m^{2})=4\pi r^{2}/(1-(v^{\varphi })^{4})$. Observe that if
the rotational velocity of the test particles were the speed of light $%
v^{\varphi }\rightarrow 1$, this area would grow very fast. Nevertheless,
for a typical galaxy, the rotational velocities are $v^{\varphi }\sim
10^{-3} $ ($300km/s$), in this case the rate of the difference of this
hypersurface area and a flat one is $(v^{\varphi })^{4}/(1-(v^{\varphi
})^{4})\sim 10^{-12}$, which is too small to be measured, but sufficient to
give the right behavior of the motion of stars in a galaxy. The effective
density depends on the velocities of the stars in the galaxy, $\rho
=(v^{\varphi })^{4}/(1-(v^{\varphi })^{4})\times 1/(\kappa _{0}r^{2})$ which
for the typical velocities in a galaxy is $\rho \sim 10^{-12}\times
1/(\kappa _{0}r^{2})$, while the effective radial pressure is $%
|P|=(v^{\varphi })^{2}((v^{\varphi })^{2}+2)/(1-(v^{\varphi })^{2})\times
1/(\kappa _{0}r^{2})\sim 10^{-6}\times 1/(\kappa _{0}r^{2})$, $i.e.$, six
orders of magnitude greater than the scalar field density. This is the
reason why it is not possible to understand a galaxy with Newtonian
dynamics. Newton theory is the limit of the Einstein theory for weak fields,
small velocities but also for small pressures (in comparison with
densities). A galaxy fulfills the first two conditions, but it has pressures
six orders of magnitude bigger than the dark matter density, which is the
dominating density in a galaxy. This effective pressure is the responsible
for the\ behavior of the flat rotation curves in the dark matter dominated
part of the galaxies.

Metric (\ref{main}) is not asymptotically flat, it could not be so. An
asymptotically flat metric behaves necessarily like a Newtonian potential
provoking that the velocity profile somewhere decays, which is not the
observed case in galaxies. Nevertheless, the energy density in the halo of
the galaxy decays as

\begin{equation}
\rho \sim \frac{10^{-12}}{\kappa _{0}r^{2}}=\frac{10^{-12}H_{0}^{-2}}{3r^{2}}%
\rho _{crit}
\end{equation}

\noindent where $H_{0}^{-1}=\sqrt{3}/h\ 10^{6}$ kpc is the Hubble parameter
and $\rho _{crit}$ is the critical density of the Universe. This means that
after a relative small distance $r_{crit}\sim \sqrt{3/h^{2}}\approx 3$ kpc
the effective density of the halo is similar as the critical density of the
Universe. One expects, of course, that the matter density around a galaxy is
smaller than the critical density, say $\rho _{around}\sim 0.0002\rho _{crit}
$, then $r_{crit}\approx 100$ kpc. Observe also that metric (\ref{main}) has
an almost flat three dimensional space-like hypersurface. The difference
between a flat three dimensional hypersurface area and the three dimensional
hypersurface area of metric (\ref{main}) is $\sim 10^{-12}$ , this is the
reason why the space-time of a galaxies seems to be so flat.

\section{The Galaxy Center}

The study of the center of the galaxy is much more complicated because we do
not have any direct observation of it. If we follow the hypothesis of the
scalar dark matter, we cannot expect that the center of a galaxy is made of
\ ``ordinary matter'' , we expect that it contains baryons, self-interacting
scalar fields, etc. There their states must in extreme conditions. At the
other hand, we think that the center of the galaxy is not static at all.
Observations suggest, for instance, an active nuclei. A regular solution
that explains with a great accuracy the galactic nucleus is the assumption
that there lies a boson star \cite{diego}, which on the other hand would be
an excellent source for the scalar dark matter \cite{bosones}. It is
possible to consider also the galactic nucleus as the current stage of an
evolving collapse, providing a boson star in the galactic center made of a
complex scalar field or a regular structure made of a time dependent real
scalar field (oscillaton), both being stable systems \cite{seidel,seidel98}.%

The energy conditions are no longer valid in nature, as it is shown by
cosmological observations on the dark energy. Why should the energy
conditions be valid in a so extreme state of matter like it is assumed to
exist in the center of galaxies? If dark matter is of scalar nature, why
should it fulfill the energy conditions in such extreme situation? In any
case, the center of the galaxy still remains as a place for speculations.%

\section{Conclusions}

We have developed most of the interesting features of a $95\%$ scalar-nature
cosmological model. The interesting implications of such a model are direct
consequences of the scalar potentials (\ref{sinh},\ref{cosh}).

The most interesting features appear in the scalar dark matter model. As we
have shown in this work, the solutions found alleviate the fine tuning
problem for dark matter. Once the scalar field $\Phi $ atarts to oscillate
around the minimum of its potential (\ref{cosh}), we can recover the
evolution of a standard cold dark matter model because the dark matter
density contrast is also recovered in the required amount. Also, we find a
Jeans lenght for this model. This provokes the suppression in the power
spectrum for small scales, that could explain the smooth core density of
galaxies and the dearth of dwarf galaxies. Up to this point, the model has
only one free parameter, $\lambda$. However, if we suppose that the scale of
suppression is $k=4.5 h \, $ Mpc, then $\lambda \approx 20.3$, and then all
parameters are completely fixed. From this, we found that $V_0 \simeq
(36.5\, {\rm eV} )^4$ and the mass of the ultra-light scalar particle is $%
m_\Phi \simeq 1.1\times 10^{-23}\, {\rm eV}$. Considering that potential ~(%
\ref{cosh}) is renormalizable, we calculated the scattering cross section $%
\sigma_{2 \rightarrow 2}$. From this, we find that the renormalization scale 
$\Lambda$ is of order of the Planck mass if we take the value required for
self-interacting dark matter, as our model is also self-interacting.

The hypothesis of the scalar dark matter is well justified at galactic level
(see also \cite{sph}) and at the cosmological level too \cite{luis2}. But at
this moment it is only that, a hypothesis which is worth to be investigated.%

Summarizing, a model for the Universe where the dark matter and energy are
of scalar nature can be realistic and could explain most of the observed
structures and features of the Universe.

\section{Acknowledgements}

We would like to thank Vladimir Avila-Reese, Michael Reisenberger, Ulises
Nucamendi and Hugo Villegas Brena for helpful discussions. L.A.U. would like
to thank Rodrigo Pelayo and Juan Carlos Arteaga Vel\'{a}zquez for many
helpful insight. F. S. G. wants to thank the criticism provided by Paolo
Salucci related to the galactic phenomenology and the ICTP for its kind
hospitality during the begining of this discussion. We also want to express
our acknowledgment to the relativity group in Jena for its kind hospitality.
This work is partly supported by CONACyT M\'exico under grant 34407-E, \ and
by grant 119259 (L.A.U.), DGAPA-UNAM IN121298 (D.N.) and by a cooperations
grant DFG-CONACyT.\\

\begin{table}[tbp]
\caption{Best-fit parameters}
\begin{tabular}{lrrrr}
\hline
Galaxy & (M/L)$_{disk}$ & (M/L)$_{bulge}$ & b & f$_0$ \\ 
&  &  & (kpc) & (kpc$^{-1}$) \\ \hline
NGC 2403 & 1.75 & -- & 1.63 & 0.0116 \\ 
& 0.04 & -- & 0.003 & $5.7\times 10^{-4}$ \\ 
NGC 2903 & 2.98 & -- & 8.33 & 0.0043 \\ 
& 0.12 & -- & 0.03 & $4.0\times 10^{-4}$ \\ 
NGC 6503 & 2.12 & -- & 1.79 & 0.013 \\ 
& 0.09 & -- & 0.01 & $8.7\times 10^{-4}$ \\ 
NGC 3198 & 2.69 & -- & 7.83 & 0.0054 \\ 
& 0.08 & -- & 0.02 & $3.0\times 10^{-4}$ \\ 
NGC 2841 & 5.39 & 3.25 & 13.85 & 0.0039 \\ 
& 0.34 & 0.36 & 0.16 & $2.5\times 10^{-4}$ \\ 
NGC 7331 & 5.06 & 1.11 & 0.845 & 0.0013 \\ 
& 0.23 & 0.06 & 0.002 & $8.9\times 10^{-5}$ \\ \hline
\end{tabular}
\end{table}

\end{multicols}
\end{document}